# On the photoionization of the outer electrons in noble gas endohedral atoms

M. Ya. Amusia[1, 2], A. S. Baltenkov[3] and L. V. Chernysheva[2]

[1]Racah Institute of Physics, the Hebrew University, Jerusalem 91904, Israel
[2]Ioffe Physical-Technical Institute, St.-Petersburg 194021, Russia
[3]Arifov Institute of Electronics, Tashkent, 700125, Uzbekistan

**Abstract**

We demonstrate the prominent modification of the outer shell photoionization cross-section in noble gas (NG) endohedral atoms $NG@F$ under the action of the fullerene F electron shell. This shell leads to two important effects, namely to strong enhancement of the cross-section due to fullerenes shell polarization under the action of the incoming electromagnetic wave and to prominent oscillation of this cross-section due to the reflection of the photoelectron from NG by the F shell.

All but He noble gas atoms are considered. The polarization of the fullerene shell is expressed via the total photoabsorption cross-section of F. The reflection of the photoelectron is taken into account in the frame of the so-called bubble potential that is a spherical $\delta$-type potential.

It is assumed in the derivations that NG is centrally located in the fullerene. It is assumed also, in accord with the existing experimental data, that the fullerenes radius $R_F$ is much bigger than the atomic radius $r_a$ and the thickness of the fullerenes shell $\Delta_F$. These assumptions permit, as it was demonstrated recently, to present the $NG@F$ photoionization cross-section as a product of the NG cross-section and two well defined calculated factors.

## 1. Introduction

In this paper we will consider the photoionization of outer shells of noble gas (NG) endohedral atoms, formed by a fullerene F inside of which a noble gas atom is staffed, NG@F. We will present data on all but He noble gases. In concrete calculations, as a fullerene F we will consider $C_{60}$.

Recently, a great deal of attention was and still is concentrated on photoionization of endohedral atoms. It was demonstrated in a number of papers [1-9] that the $C_{60}$ shell adds prominent resonance structure in the photoionization cross section of endohedral atoms. Although the experimental investigation of $A@C_{60}$ photoionization seems to be at this moment very difficult, it will be inevitably intensively studied in the nearest future[1]. This justifies the current efforts of the theorists that are predicting rather non-trivial effects waiting for verification.

The role of $C_{60}$ in $NG@C_{60}$ photoionization is manifold. $C_{60}$ act as a spherical potential resonator that reflects the photoelectron wave coming from NG atom. This leads to interference of out-going and in-coming (reflected) waves and to confinement resonances or simply to oscillations in the frequency dependence of the photoionization cross sections [5]. The interference of the photoelectron spherical waves inside the resonator $C_{60}$ affects significantly not only the total cross section but also the angular distribution of photoelectrons. This phenomenon was analysed in [6] where it was shown that the effects of confinement resonances are found also in the frequency dependencies of the dipole and nondipole parameters of the photoelectron angular distribution. The results of these studies give evidence that the reflection and refraction of the photoelectron waves by the potential resonator $C_{60}$ is prominent up to 60 – 80 eV of the photoelectron energy.

---

[1] As a first example of such a research, let us mention the tentative data on measurements of photoionization cross-section of $Ce@C_{82}$ [10].

The $C_{60}$ shell at some frequencies acts as a dynamical screen that is capable to suppress or enhance the incident electromagnetic radiation acting upon the doped atom A [11-13]. This effect is due to dynamical polarization of the collectivized electrons of the fullerene shell. Plasma excitations of these electrons generate an alternating dipole moment. This dipole moment causes the ionization of the electronic shells of the endohedral atom. The screening effects of the $C_{60}$ shell are particularly strong for incident radiation frequency $\omega$[2] of about that of the $C_{60}$ Giant resonance, i.e. 20 – 22 eV, but is big enough in a much broader region, from ionization threshold up to 60 – 80 eV.

We will show in this paper that the dynamic polarization of $C_{60}$ increases the outer shell photoionization cross-section at any $\omega$, contrary to the statements in [11]. The maximal enhancement is in the region of $C_{60}$ dipole polarizability maximum.

Thus, the resonator and dynamic screen effects of the fullerene shell $C_{60}$ manifest themselves as a prominent enhancement of the cross-section modulated by oscillating structure that appears due to reflection and refraction of the photoelectron wave by the fullerenes shell.

We have studied before [12, 14] the effect of $C_{60}$ upon the most important atomic resonances, the Giant and interference, both in Xe, and found prominent modifications of them. However the modification of the outer shell was out of our attention, in spite of the fact that by far the biggest part of the absolute cross-section comes from the near threshold region of the outer subshell. This minus of all previous considerations is eliminated in this paper.

It would be of interest to see the alteration of the photoionization cross-section if instead of $C_{60}$ other fullerenes, like $C_{70}$, $C_{76}$, $C_{82}$ or $C_{87}$ are considered. However, in order to study the endohedrals NG@F with F = $C_{70}$, $C_{76}$, $C_{82}$ or $C_{87}$ one needs to know the shape of these objects, their photoionization cross-sections and the position of the NG atoms inside the fullerenes cage. The answers to these questions are absent at this moment.

Generally speaking, there is no doubt about the existence of more or less pronounced oscillations due to the photoelectron reflection by any fullerene F shell. However the possibility of enhancement due to fullerenes shell dynamic polarization is considerably more problematic.

**2. Main formulas**

We will use here the theoretical approaches already developed in a number of previous papers [12-14]. However, for completeness, let us repeat the main points of the consideration and present the essential formula used in calculations.

---

[2] Atomic system of units is used in this paper

Let us start with the problem of an isolated closed shell atom. The following relation gives the differential in angle photoionization cross-section by non-polarized light of frequency $\omega$ [15, 16]:

$$\frac{d\sigma_{nl}(\omega)}{d\Omega} = \frac{\sigma_{nl}(\omega)}{4\pi}[1 - \frac{\beta_{nl}}{2}P_2(\cos\theta) + \kappa\gamma_{nl}P_1(\cos\theta) + \kappa\eta_{nl}P_3(\cos\theta)], \quad (1)$$

where $\kappa = \omega/c$, $P_l(\cos\theta)$ are the Legendre polynomials, $\theta$ is the angle between photon momentum $\mathbf{\kappa}$ and photoelectron velocity $\mathbf{v}$, $\beta_{nl}(\omega)$ is the dipole, while $\gamma_{nl}(\omega)$ and $\eta_{nl}(\omega)$ are so-called non-dipole angular anisotropy parameters.

Since in experiment, usually the sources of linearly polarized radiation are used, instead of (1) another form of angular distribution is more convenient [17, 18]:

$$\frac{d\sigma_{nl}(\omega)}{d\Omega} = \frac{\sigma_{nl}(\omega)}{4\pi}[1 + \beta_{nl}P_2(\cos\Theta) + (\delta_{nl}^C + \gamma_{nl}^C \cos^2\Theta)\sin\Theta\cos\Phi]. \quad (2)$$

Here $\Theta$ is the polar angle between the vectors of photoelectron's velocity $\mathbf{v}$ and photon's polarization $\mathbf{e}$, while $\Phi$ is the azimuth angle determined by the projection of $\mathbf{v}$ in the plane orthogonal to $\mathbf{e}$ that includes the vector of photon's velocity. The non-dipole parameters in (1) and (2) are connected by the simple relations [17]

$$\frac{\gamma_{nl}^C}{5} + \delta_{nl}^C = \kappa\gamma_{nl}, \qquad \frac{\gamma_{nl}^C}{5} = -\kappa\eta_{nl}. \quad (3)$$

The below-presented concrete results of calculations of non-dipole parameters are obtained using expression (2). There are two possible dipole transitions from subshell $l$, namely $l \to l\pm 1$ and three quadrupole transitions $l \to l; l\pm 2$. Corresponding general expressions for $\beta_{nl}(\omega)$, $\gamma_{nl}(\omega)$ and $\eta_{nl}(\omega)$ are rather complex and are presented as combinations of dipole $d_{l\pm 1}$ and quadrupole $q_{l\pm 2,0}$ matrix elements of photoelectron transitions and photoelectrons waves phases. In one-electron Hartree-Fock (HF) approximation these parameters are presented as [19]:

$$\beta_{nl}(\omega) = \frac{1}{(2l+1)[(l+1)d_{l+1}^2 + ld_{l-1}^2]}[(l+1)(l+2)d_{l+1}^2 + l(l-1)d_{l-1}^2 -$$
$$6l(l+1)d_{l+1}d_{l-1}\cos(\delta_{l+1} - \delta_{l-1})]. \quad (4)$$

As to the parameters $\gamma_{nl}(\omega)$, $\eta_{nl}(\omega)$, they are given by the following expressions [16]:

$$\gamma_{nl}(\omega) = \frac{3}{5[ld_{l-1}^2 + (l+1)d_{l+1}^2]}\left\{\frac{l+1}{2l+3}[3(l+2)q_{l+2}d_{l+1}\cos(\delta_{l+2} - \delta_{l+1}) - lq_l d_{l+1} \times \right.$$
$$\left. \times\cos(\delta_{l+2} - \delta_{l+1})] - \frac{l}{2l+1}[3(l-1)q_{l-2}d_{l-1}\cos(\delta_{l-2} - \delta_{l-1}) - (l+1)q_l d_{l-1}\cos(\delta_l - \delta_{l-1})]\right\}, \quad (5)$$

$$\eta_{nl}(\omega) = \frac{3}{5[ld_{l-1}^2 + (l+1)d_{l+1}^2]} \left\{ \frac{(l+1)(l+2)}{(2l+1)(2l+3)} q_{l+2}[5ld_{l-1}\cos(\delta_{l+2} - \delta_{l-1}) - \right.$$
$$- (l+3)d_{l+1}\cos(\delta_{l+2} - \delta_{l-1})] - \frac{(l-1)l}{(2l+1)(2l+1)} q_{l-2} \times$$
$$\times [5(l+1)d_{l+1}\cos(\delta_{l-2} - \delta_{l+1}) - (l-2)d_{l-1}\cos(\delta_{l-2} - \delta_{l-1})] +$$
$$+ 2\frac{l(l+1)}{(2l-1)(2l+3)} q_l [(l+2)d_{l+1}\cos(\delta_l - \delta_{l+1}) - (l-1)d_{l-1}\cos(\delta_l - \delta_{l-1})] \right\}. \quad (6)$$

Here $\delta_l(k)$ are the photoelectrons' scattering phases; the following relation gives the matrix elements $d_{l\pm 1\uparrow\downarrow}$ in the so-called *r*-form

$$d_{l\pm 1} \equiv \int_0^\infty P_{nl}(r) r P_{\varepsilon l\pm 1}(r) dr, \quad (7)$$

where $P_{nl}(r)$, $P_{\varepsilon l\pm 1}(r)$ are the radial Hartree-Fock (HF) [19] one-electron wave functions of the *nl* discrete level and $\varepsilon l \pm 1$ - in continuous spectrum, respectively. The following relation gives the quadrupole matrix elements

$$q_{l\pm 2,0} \equiv \frac{1}{2} \int_0^\infty P_{nl}(r) r^2 P_{\varepsilon l\pm 2,0}(r) dr. \quad (8)$$

In order to take into account the Random Phase Approximation with Exchange (RPAE) [19] multi-electron correlations, one has to perform the following substitutions in the expressions for $\beta_{nl}(\omega)$, $\gamma_{nl}(\omega)$ and $\eta_{nl}(\omega)$ [16]:

$$d_{l+1}d_{l-1}\cos(\delta_{l+1} - \delta_{l-1}) \to [(\text{Re}\,D_{l+1}\,\text{Re}\,D_{l-1} + \text{Im}\,D_{l+1}\,\text{Im}\,D_{l\pm 2,0})\cos(\delta_{l\pm 2,0} - \delta_{l\pm 1}) -$$
$$- (\text{Re}\,D_{l+1}\,\text{Im}\,D_{l-1} - \text{Im}\,D_{l+1}\,\text{Re}\,Q_{l-1})\sin(\delta_{l+1} - \delta_{l-1})] \equiv \quad (9)$$
$$\equiv \tilde{D}_{l+1}\tilde{D}_{l-1}\cos(\delta_{l\pm 2,0} + \Delta_{l\pm 2,0} - \delta_{l\pm 1} - \Delta_{l\pm 1}).$$

$$d_{l\pm 1}q_{l\pm 2,0}\cos(\delta_{l\pm 2,0} - \delta_{l\pm 1}) \to [(\text{Re}\,D_{l\pm 1}\,\text{Re}\,Q_{l\pm 2,0} + \text{Im}\,D_{l\pm 1}\,\text{Im}\,Q_{l\pm 2,0})\cos(\delta_{l\pm 2,0} - \delta_{l\pm 1}) -$$
$$- (\text{Re}\,D_{l\pm 1}\,\text{Im}\,Q_{l\pm 2,0} - \text{Im}\,D_{l\pm 1}\,\text{Re}\,Q_{l\pm 2,0})\sin(\delta_{l\pm 2,0} - \delta_{l\pm 1})] \equiv \quad (10)$$
$$\equiv \tilde{D}_{l\pm 1}\tilde{Q}_{l\pm 2,0}\cos(\delta_{l\pm 2,0} + \Delta_{l\pm 2,0} - \delta_{l\pm 1} - \Delta_{l\pm 1}),$$
$$d_{l\pm 1}^2 \to \text{Re}\,D_{l\pm 1}^2 + \text{Im}\,D_{l\pm 1}^2 \equiv \tilde{D}_{l\pm 1}^2.$$

Here the following notations are used for the matrix elements with account of multi-electron correlations, dipole and quadrupole, respectively:

$$D_{l\pm 1}(\omega) \equiv \tilde{D}_{l\pm 1}(\omega)\exp[i\Delta_{l\pm 1}(\varepsilon)], \quad Q_{l\pm 2,0}(\omega) \equiv \tilde{Q}_{l\pm 2,0}(\omega)\exp[i\Delta_{l\pm 2,0}(\varepsilon)], \quad (11)$$

where $\tilde{D}_{l\pm 1}(\omega)$, $\tilde{Q}_{l\pm 2,0}(\omega)$, $\Delta_{l\pm 1}$ and $\Delta_{l\pm 2,0}$ are absolute values of the amplitudes for respective transitions and phases for photoelectrons with angular moments $l \pm 1$ and $l \pm 2,0$.

The following are the RPAE equation for the dipole matrix elements [19]

$$\langle v_2|D(\omega)|v_1\rangle = \langle v_2|d|v_1\rangle + \sum_{v_3,v_4} \frac{\langle v_3|D(\omega)|v_4\rangle(n_{v_4}-n_{v_3})\langle v_4 v_2|U|v_3 v_1\rangle}{\varepsilon_{v_4}-\varepsilon_{v_3}+\omega+i\eta(1-2n_{v_3})}, \qquad (12)$$

where

$$\langle v_1 v_2|\hat{U}|v_1' v_2'\rangle \equiv \langle v_1 v_2|\hat{V}|v_1' v_2'\rangle - \langle v_1 v_2|\hat{V}|v_2' v_1'\rangle. \qquad (13)$$

Here $\hat{V} \equiv 1/|\vec{r}-\vec{r}'|$ and $v_i$ is the total set of quantum numbers that characterize a HF one-electron state on discrete (continuum) levels. That includes the principal quantum number (energy), angular momentum, its projection and the projection of the electron spin. The function $n_{v_i}$ (the so-called step-function) is equal to 1 for occupied and 0 for vacant states.

The dipole matrix elements $D_{l\pm 1}$ are obtained by solving the radial part of the RPAE equation (12). As to the quadrupole matrix elements $Q_{l\pm 2,0}$, they are obtained by solving the radial part of the RPAE equation, similar to (12)

$$\langle v_2|Q(\omega)|v_1\rangle = \langle v_2|\hat{q}|v_1\rangle + \sum_{v_3,v_4} \frac{\langle v_3|Q(\omega)|v_4\rangle(n_{v_4}-n_{v_3})\langle v_4 v_2|U|v_3 v_1\rangle}{\varepsilon_{v_4}-\varepsilon_{v_3}+\omega+i\eta(1-2n_{v_3})}. \qquad (14)$$

Here in $r$-form one has $\hat{q} = r^2 P_2(\cos\theta)$.
Equations (12, 14) are solved numerically using the procedure discussed at length in [20].

## 3. Effect of C$_{60}$ fullerene shell

Let us start with the confinement effects. These effects near the photoionization threshold can be described within the framework of the "orange" skin potential model. According to this model, for small photoelectron energies the real static potential of the C$_{60}$ can be presented by the zero-thickness bubble pseudo-potential (see [21, 22] and references therein):

$$V(r) = -V_0 \delta(r-R). \qquad (15)$$

The parameter $V_0$ is determined by the requirement that the binding energy of the extra electron in the negative ion $C_{60}^-$ is equal to its observable value. Addition of the potential (15) to the atomic HF potential leads to a factor $F_l(k)$ in the photoionization amplitudes, which depends only upon the photoelectron's momentum $k$ and orbital quantum number $l$ [21, 22]:

$$F_l(k) = \cos\breve{\Delta}_l(k)\left[1 - \tan\breve{\Delta}_l(k)\frac{v_{kl}(R)}{u_{kl}(R)}\right], \qquad (16)$$

where $\breve{\Delta}_l(k)$ are the additional phase shifts due to the fullerene shell potential (15). They are expressed by the following formula:

$$\tan \breve{\Delta}_l(k) = \frac{u_{kl}^2(R)}{u_{kl}(R)v_{kl}(R) + k/2V_0} . \tag{17}$$

In these formulas $u_{kl}(r)$ and $v_{kl}(r)$ are the regular and irregular solutions of the atomic HF equations for a photoelectron with momentum $k = \sqrt{2\varepsilon}$, where $\varepsilon$ is the photoelectron energy connected with the photon energy $\omega$ by the relation $\varepsilon = \omega - I_A$ with $I$ being the atom A ionization potential.

Using Eq. (16), one can obtain the following relation for $D^{AC(r)}$ and $Q^{AC(r)}$ amplitudes for endohedral atom A@C$_{60}$ with account of photoelectron's reflection and refraction by the C$_{60}$ static potential (15), expressed via the respective values for isolated atom that correspond to $nl \to \varepsilon l'$ transitions:

$$\begin{aligned} D^{AC(r)}_{nl,kl'}(\omega) &= F_{l'}(k) D_{nl,kl'}(\omega), \\ Q^{AC(r)}_{nl,kl''}(\omega) &= F_{l''}(k) Q_{nl,kl''}(\omega) \end{aligned} . \tag{18}$$

For the cross-sections one has

$$\sigma^{AC(r)}_{nl,kl'}(\omega) = [F_{l'}(k)]^2 \sigma^A_{nl,kl'}(\omega), \tag{19}$$

where $\sigma^A_{nl,kl'}(\omega)$ is the contribution of the $nl \to \varepsilon l'$ transition to the photoionization cross-section of atomic subshell $nl$, $\sigma^A_{nl}(\omega)$.

Now let us discuss the role of polarization of the C$_{60}$ shell under the action of the photon beam [12]. The effect of the fullerene electron shell polarization upon atomic photoionization amplitude can be taken into account in RPAE using (12). An essential simplification comes from the fact, that the C$_{60}$ radius $R_C$ is much bigger than the atomic radius $r_a$, $R_C \gg r_a$. It is also important that the electrons in C$_{60}$ are located within a layer, the thickness of which $\Delta_C$ is considerably smaller than $R_C$. In this case, the amplitude of endohedral atom's photoionization due to $nl \to \varepsilon l'$ transition with all essential atomic correlations taken into account can be presented by the following formula [12]:

$$D^{AC}_{nl,\varepsilon l'}(\omega) \cong F_{l'}(k)\left(1 - \frac{\alpha^d_C(\omega)}{R_C^3}\right) D^A_{nl,\varepsilon l'}(\omega) \equiv F_{l'}(k) G^d(\omega) D^A_{nl,\varepsilon l'}(\omega), \tag{20}$$

where $\alpha^d_C(\omega)$ is the dipole dynamical polarizability of C$_{60}$ and $R_C$ is its fullerenes radius. For the quadrupole amplitude one obtains a similar expression:

$$Q^{AC}_{nl,\varepsilon l'}(\omega) \cong F_{l'}(k)\left(1 - \frac{\alpha^q_C(\omega)}{R_C^5}\right) Q^A_{nl,\varepsilon l'}(\omega) \equiv F_{l'}(k) G^q(\omega) Q^A_{nl,\varepsilon l'}(\omega), \tag{21}$$

where $\alpha_C^q(\omega)$ is the quadrupole dynamical polarizability of $C_{60}$. The $G^{d,q}(\omega)$ factors are complex numbers that we present as

$$G^{d,q}(\omega) = \tilde{G}^{d,q}(\omega)\exp[i\eta^{d,q}(\omega)], \qquad (22)$$

where $\tilde{G}^{d,q}(\omega)$ are respective absolute values.

Using the relation between the imaginary part of the polarizability and the dipole photoabsorption cross-section $\sigma_C^d(\omega)$ - $\text{Im}\,\alpha_C^d(\omega) = c\sigma_C^d(\omega)/4\pi\omega$, one can derive the polarizability of the $C_{60}$ shell. Although experiments [23, 24] do not provide absolute values of $\sigma_C^d(\omega)$, it can be reliably estimated using different normalization procedures on the basis of the sun rule: $(c/2\pi^2)\int_{I_o}^{\infty}\sigma_C^d(\omega)d\omega = N$, where $N$ is the number of collectivized electrons. The real part of polarizability is connected with imaginary one (and with the photoabsorption cross-section) by the dispersion relation:

$$\text{Re}\,\sigma_C^d(\omega) = \frac{c}{2\pi^2}\int_{I_C}^{\infty}\frac{\sigma_C^d(\omega')d\omega'}{\omega'^2 - \omega^2}, \qquad (23)$$

where $I_C$ is the $C_{60}$ ionization potential. This approach was used for polarizability of $C_{60}$ in [13], where it was considered that $N = 240$, i.e. 4 collectivized electrons per each C atom in $C_{60}$. Using the photoabsorption data that are considered as most reliable in [23], we obtained $N_{eff} \approx 250$ that is sufficiently close to the value, assumed in [13].

The equality $\text{Im}\,\alpha_C^q(\omega) = c\sigma_C^q(\omega)/4\pi\omega$ and quadrupole dispersion relation similar to (24) are valid. But the quadrupole photoabsorption cross-section is so small that it cannot be derived experimentally.

Note that because the strong inequality $R_C \gg r_a$ we have derived formulas (20) and (21) that are more accurate than those obtained from the RPAE for the whole A@F system. This is important since "one electron – one vacancy" channel that is the only taken into account in RPAE is not always dominant in the photoabsorption cross-section of the fullerene and hence in its polarizability.

Using the amplitude (20), one has for the cross section

$$\sigma_{nl,\varepsilon l'}^{AC}(\omega) = [F_{l'}(\omega)]^2\left|1 - \frac{\alpha_C^d(\omega)}{R_C^3}\right|^2\sigma_{nl,\varepsilon l'}^A(\omega) \equiv [F_{l'}(\omega)]^2 S(\omega)\sigma_{nl,\varepsilon l'}^A(\omega), \qquad (24)$$

where $S(\omega) = [\tilde{G}^d(\omega)]^2$ cab be called radiation enhancement parameter.

With these amplitudes, using the expressions (4-6) and performing the substitution (9, 10) we obtain the cross-sections for ND@$C_{60}$ and angular anisotropy parameters. While calculating the anisotropy parameter, the cosines of atomic phases differences $\cos(\delta_l - \delta_{l'})$ in formulas (4)-(6) are replaced at first by $\cos(\delta_l + \Delta_l - \delta_{l'} - \Delta_{l'})$. As a result, one has for the dipole angular anisotropy parameter (4), using (9):

$$\beta_{nl}(\omega) = \frac{1}{(2l+1)\left[(l+1)F_{l+1}^2\tilde{D}_{l+1}^2 + lF_{l-1}^2\tilde{D}_{l-1}^2\right]}[(l+1)(l+2)F_{l+1}^2\tilde{D}_{l+1}^2$$
$$+ l(l-1)F_{l-1}^2\tilde{D}_{l-1}^2 - 6l(l+1)F_{l+1}F_{l-1}\tilde{D}_{l+1}\tilde{D}_{l-1}\cos(\tilde{\delta}_{l+1} - \tilde{\delta}_{l-1})] \qquad . \qquad (25)$$

where $\tilde{\delta}_{l'} = \delta_{l'} + \Delta_{l'}$ (see (11) ). Naturally, the dipole parameter $\beta_{nl}(\omega)$ is not affected by $G^d(\omega)$ factors that similarly alter the nominator and denominator in (25).

The situation for non-dipole parameters is different, since $G^d(\omega) \neq G^q(\omega)$. From (5) and (6), using (9) and (10) we arrive to the following expressions for the non-dipole angular anisotropy parameters:

$$\gamma_{nl}(\omega) = \frac{3\tilde{G}^q(\omega)}{5\tilde{G}^d(\omega)\left[(l+1)F_{l+1}^2\tilde{D}_{l+1}^2 + lF_{l-1}^2\tilde{D}_{l-1}^2\right]} \times$$
$$\times \left\{ \frac{(l+1)F_{l+1}}{2l+3}[3(l+2)F_{l+2}\tilde{Q}_{l+2}\tilde{D}_{l+1}\cos\left(\tilde{\tilde{\delta}}_{l+2} - \tilde{\tilde{\delta}}_{l+1}\right) - lF_l\tilde{Q}_l\tilde{D}_{l+1}\cos\left(\tilde{\tilde{\delta}}_{l+2} - \tilde{\tilde{\delta}}_{l+1}\right)] - \right. \qquad (26)$$
$$\left. - \frac{lF_{l-1}}{2l+1}\left[3(l-1)F_{l-2}\tilde{Q}_{l-2}\tilde{D}_{l-1}\cos\left(\tilde{\tilde{\delta}}_{l-2} - \tilde{\tilde{\delta}}_{l-1}\right) - (l+1)F_l\tilde{Q}_l\tilde{D}_{l-1}\cos\left(\tilde{\tilde{\delta}}_l - \tilde{\tilde{\delta}}_{l-1}\right)\right] \right\},$$

$$\eta_{nl}(\omega) = \frac{3\tilde{G}^q(\omega)}{5\tilde{G}^d(\omega)\left[(l+1)F_{l+1}^2\tilde{D}_{l+1}^2 + lF_{l-1}^2\tilde{D}_{l-1}^2\right]} \times$$
$$\times \left\{ \frac{(l+1)(l+2)}{(2l+1)(2l+3)}F_{l+2}\tilde{Q}_{l+2}\left[5lF_{l-1}\tilde{D}_{l-1}d_{l-1}\cos\left(\tilde{\tilde{\delta}}_{l+2} - \tilde{\tilde{\delta}}_{l-1}\right) - (l+3)F_{l+1}\tilde{D}_{l+1}\cos\left(\tilde{\tilde{\delta}}_{l+2} - \tilde{\tilde{\delta}}_{l-1}\right)\right] - \right.$$
$$- \frac{(l-1)l}{(2l+1)(2l+1)}F_{l-2}\tilde{Q}_{l-2}\left[5(l+1)F_{l+1}\tilde{D}_{l+1}\cos\left(\tilde{\tilde{\delta}}_{l-2} - \tilde{\tilde{\delta}}_{l+1}\right) - (l-2)\tilde{F}_{l-1}\tilde{D}_{l-1}\cos\left(\tilde{\tilde{\delta}}_{l-2} - \tilde{\tilde{\delta}}_{l-1}\right)\right] +$$
$$\left. + 2\frac{l(l+1)F_l\tilde{Q}_l}{(2l-1)(2l+3)}\left[(l+2)F_{l+1}\tilde{D}_{l+1}\tilde{D}_{l+1}\cos\left(\tilde{\tilde{\delta}}_l - \tilde{\tilde{\delta}}_{l+1}\right) - (l-1)F_{l-1}\tilde{D}_{l-1}\tilde{D}_{l-1}\cos\left(\tilde{\tilde{\delta}}_l - \tilde{\tilde{\delta}}_{l-1}\right)\right] \right\},$$
(27)

where $\tilde{\tilde{\delta}}_{l\pm2,0} = \tilde{\delta}_{l\pm2,0} + \eta^q$ and $\tilde{\tilde{\delta}}_{l\pm1} = \tilde{\delta}_{l\pm1} + \eta^d$ (see (22)).

## 4. Some details of calculations and their results

Naturally, the $C_{60}$ parameters in the present calculations were chosen the same as in the previous papers, e.g. in [22]: $R = 6.639$ and $V_0 = 0.443$.

Preliminary investigations have demonstrated that $G_C^q(\omega)$ is close to unity. That means that the role of quadrupole polarization can be neglected. This is why in the below-presented results we assume that $G_C^q(\omega) = 1$ and $\eta^q = 0$. We plan to pay special attention to the role of quadrupole excitations in general and quadrupole continuous spectrum resonances in particular in the future. However without experimental data this task is far from being trivial.

Previously we have presented the cross-section and dipole angular anisotropy parameter for 5p electrons in Xe@$C_{60}$ [25]. Here, in Fig2-9 we have in addition data for the outer shell photoionization cross-section as well as dipole and non-dipole angular anisotropy parameters of outer p-electrons in NG@$C_{60}$, where NG=Ne, Ar, Kr, Xe.

In Fig. 1 we present the radiation enhancement parameter $S(\omega)$, its amplitude's absolute value $\tilde{G}^d(\omega) \equiv |G(\omega)|$ and phase $\eta^d \equiv \arg G(\omega)$. It is seen that the curves are rather complex and able modify considerably the outer subshell photoionization cross-section and non-dipole angular anisotropy parameters.

Fig. 2 depicts the photoionization cross-section and dipole angular anisotropy parameters for $2p$ electrons of Ne@$C_{60}$, while Fig. 3 presents the data for non-dipole parameters. In a similar way the same characteristics are presented in Fig. 4-9 for Ar@$C_{60}$, Kr@$C_{60}$, and Xe@$C_{60}$.

In all considered cases we see prominent influence of the fullerenes shell upon the photoionization of the "caged" atom – Ne, Ar, Kr and Xe. Most impressive is the increase of the photoionization cross-section that reaches a factor of fifteen – twenty. The absolute value of the cross-section reaches in atomic scale tremendous values, up to 1000 Mb.

The maxima in upper curves in Fig. 2, 4, 6, 8 are reasonable to call Giant Endohedral resonance - GER. No doubt that similar effect takes place in outer shells of other endohedral atoms and for atoms "caged" by other than $C_{60}$ fullerenes. It is quite probable that such resonances can be detected in experimental studies of photoionization of endohedral atoms, using the photoelectron spectroscopy methods.

The GER in photoionization of Ar@$C_{60}$, Kr@$C_{60}$ and Xe@$C_{60}$ impressively exceeds the atomic Giant resonance in the photoionization cross-section of Xe $4d$ electrons. The monotonically decreasing curves in isolated atoms are transformed into curves with two maxima (one big and the other much smaller) with a remarkably big total oscillator strength of about 25, i.e. 2.5 times bigger than that of the $4d$ atomic Giant resonance in isolated Xe. Due to this increase of the "caged" atom photoionization cross-section, the total oscillator strength sum in the $\omega$ range, say, $I < \omega < I + 1Ry$ increases dramatically.

A natural question is: what is the origin of this increase? The answer is as follows – this increase comes from the fullerene shell, and not due to redistribution of the "caged" atom oscillator strength. The latter is evident, since this atom, simply speaking, has not enough electrons for this. Note that the total sum rule for an endohedral is equal to $N_F + N_A$, where $N_F$ and $N_A$ are the total numbers of electrons in the fullerene and the atom, respectively. As to the sum of oscillator strengths of the "caged" atom, it increases, roughly speaking, by the area on the Fig. 2, 4, 6, 8 that is between the solid and dashed-dotted curves.

The effect of photoelectrons' reflection upon dipole angular anisotropy parameter $\beta$ is illustrated by the lover curves in Fig. 2, 4, 6, 8. It is seen that fullerenes shell adds oscillations that are particularly strong in Ne and Xe.

Non-dipole parameters are depicted in Fig. 3, 5, 7, 9 for Ne, Ar, Kr, and Xe, respectively. Here the reflection of the photoelectron by the fullerenes shell lead to oscillations, while the polarization of the shell results for $\gamma^C$ in prominent decrease. This parameter is weakly dependent upon the phase $\eta$. As to $\delta^C$, although being much smaller than $\gamma^C$, it is considerably more sensitive to $\eta$. As a result, $\delta^C$ becomes not only smaller but even bigger, as is seen in Fig. 9, in spite of the fact that it has $G(\omega) > 1$ in the denominator.

One has to have in mind that in reality $C_{60}$ is a non-spherical structure. This can noticeably affect the experimentally measured angular anisotropy parameters of endohedral atoms. The finite thickness of the fullerene shell that is neglected by our model potential (15) can to some extent smooth out the predicted oscillations in the cross-sections and anisotropy parameters. However, other theoretical approaches, being much more complex than ours (see e.g. [1]) also suffer from essential shortcomings. Further progress in this area needs experimental data at least on the endohedral atoms photoionization cross-sections.

**Acknowledgement**

MYaA is grateful for financial support to the Israeli Science Foundation, Grant 174/03 and the Hebrew University Intramural Funds. ASB expresses his gratitude to the Hebrew University for hospitality and for financial support by Uzbekistan National Foundation, Grant Ф-2-1-12.

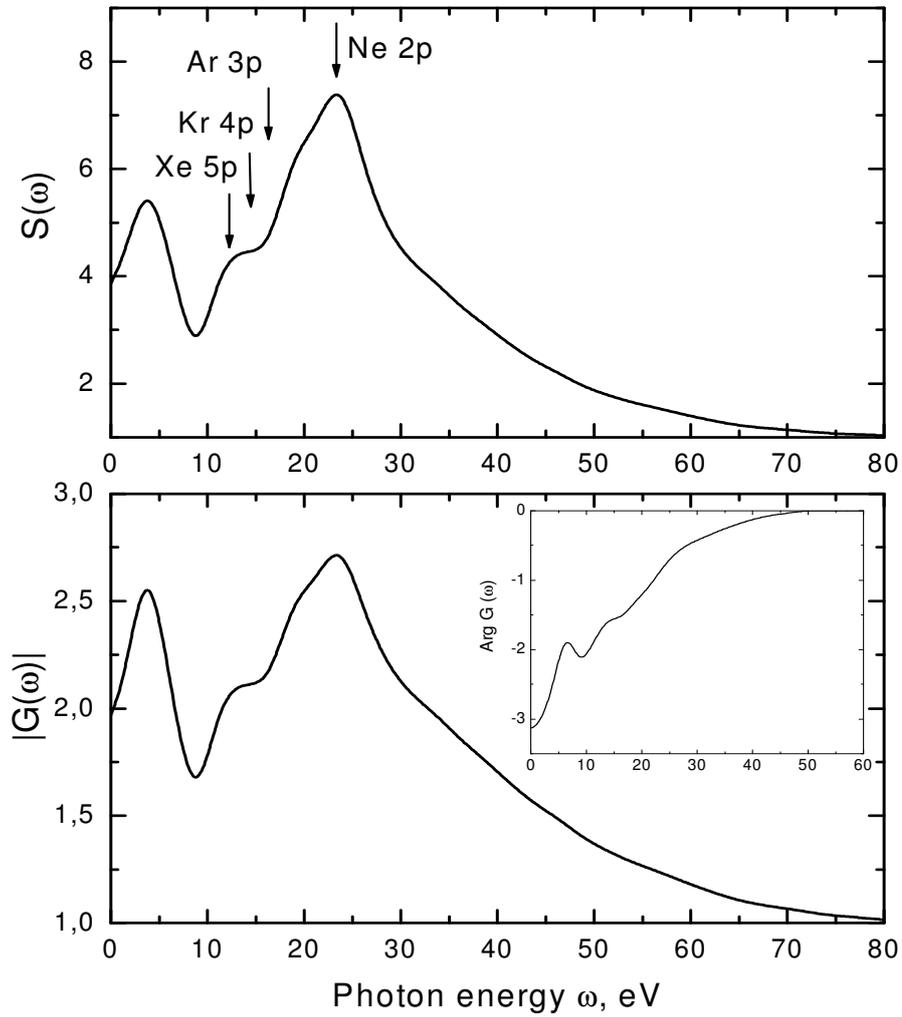

Fig.1. Radiation enhancement parameter $S(\omega)$, its amplitude's absolute value $\tilde{G}^d(\omega) \equiv |G(\omega)|$ and phase $\eta^d \equiv \arg G(\omega)$. Arrows denote the thresholds positions of corresponding outer $np$ subshells.

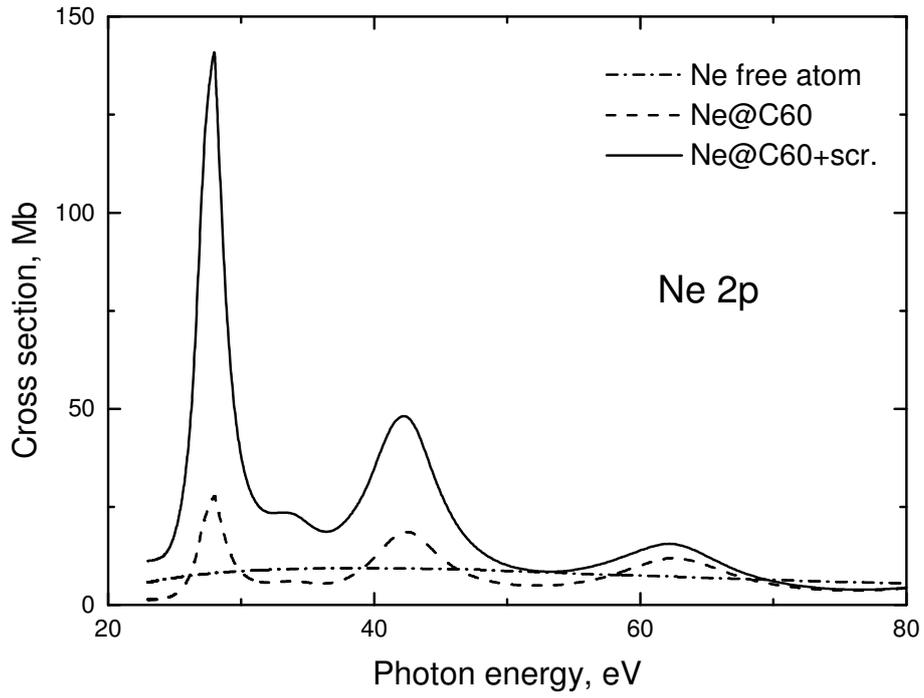

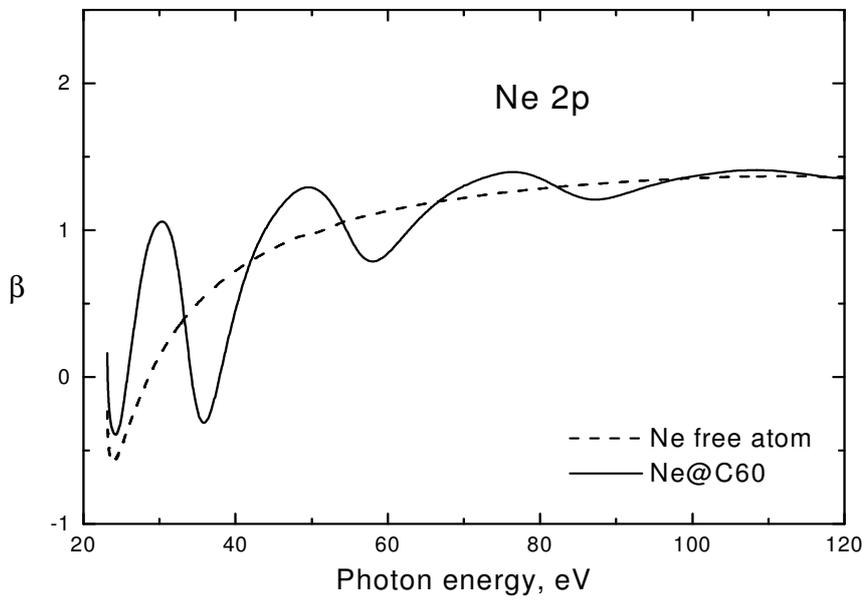

Fig. 2. Photoionization cross-section and dipole angular anisotropy parameter $\beta$ for 2p electrons of Ne@$C_{60}$

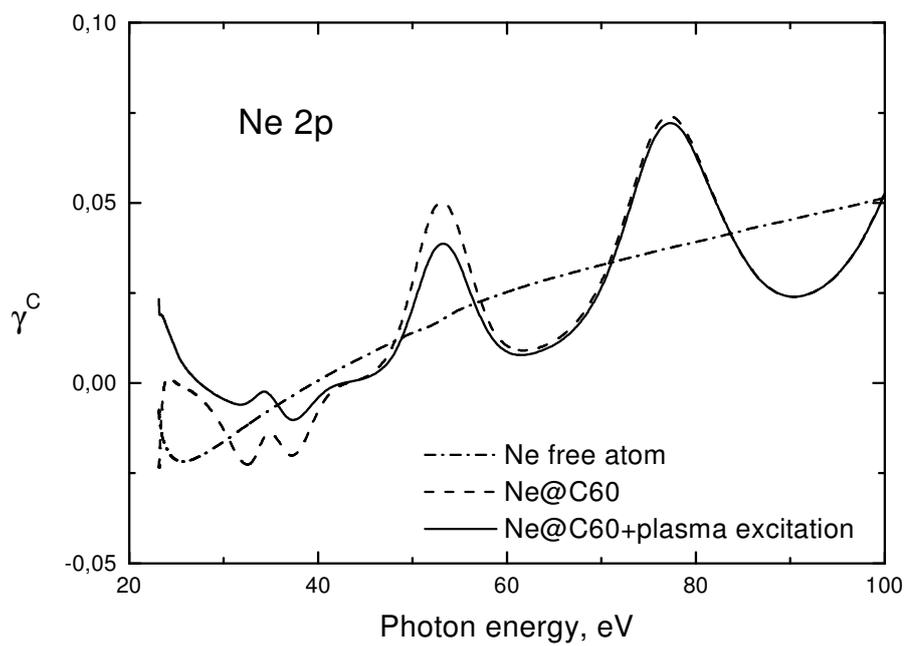

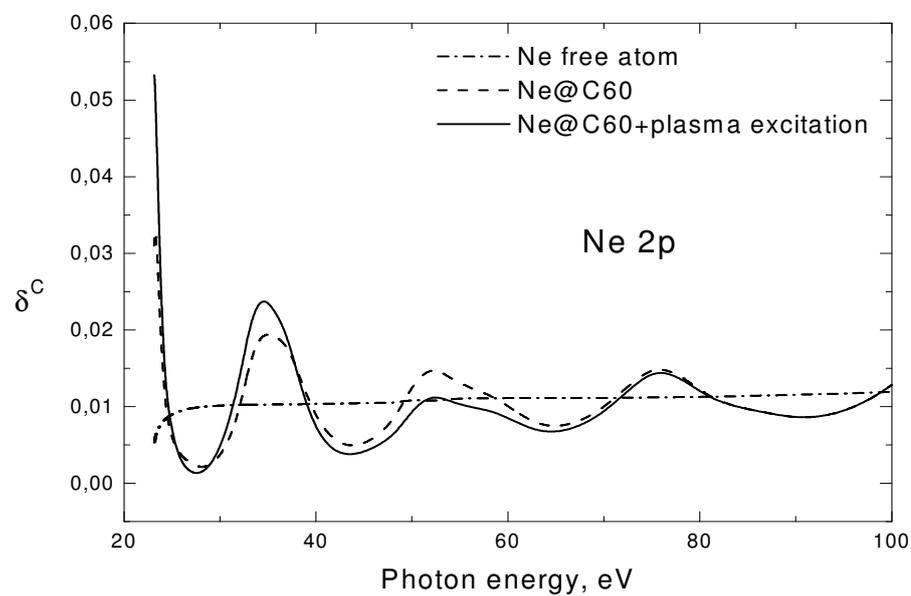

Fig.3. Non-dipole angular anisotropy parameters $\gamma^C$ and $\delta^C$ for 2p electrons of Ne@C$_{60}$

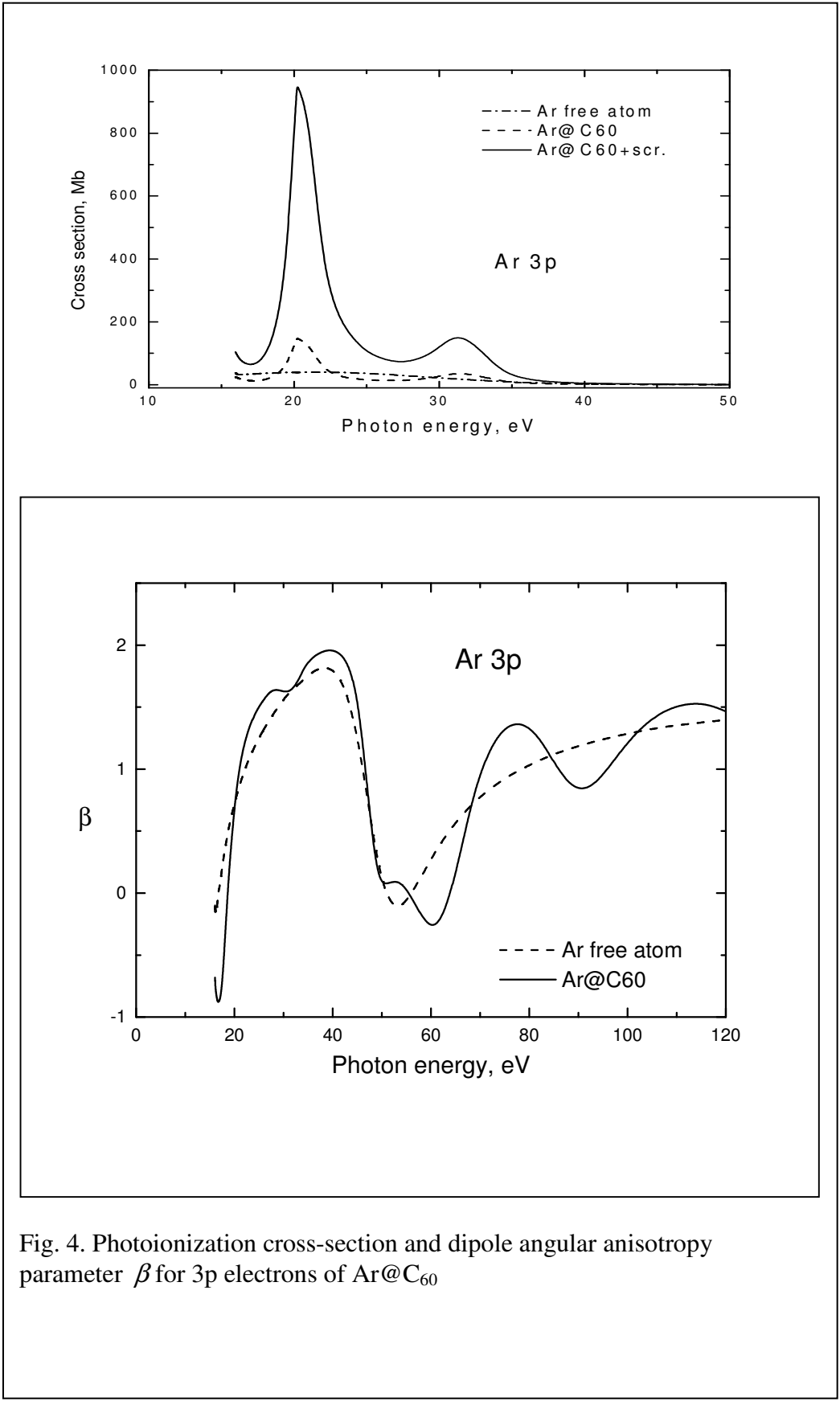

Fig. 4. Photoionization cross-section and dipole angular anisotropy parameter $\beta$ for 3p electrons of Ar@$C_{60}$

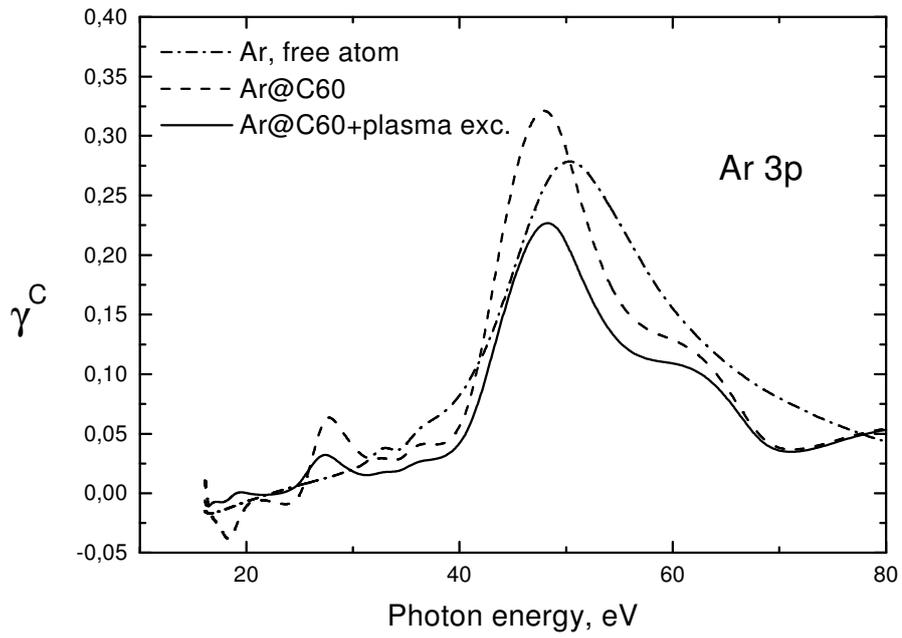

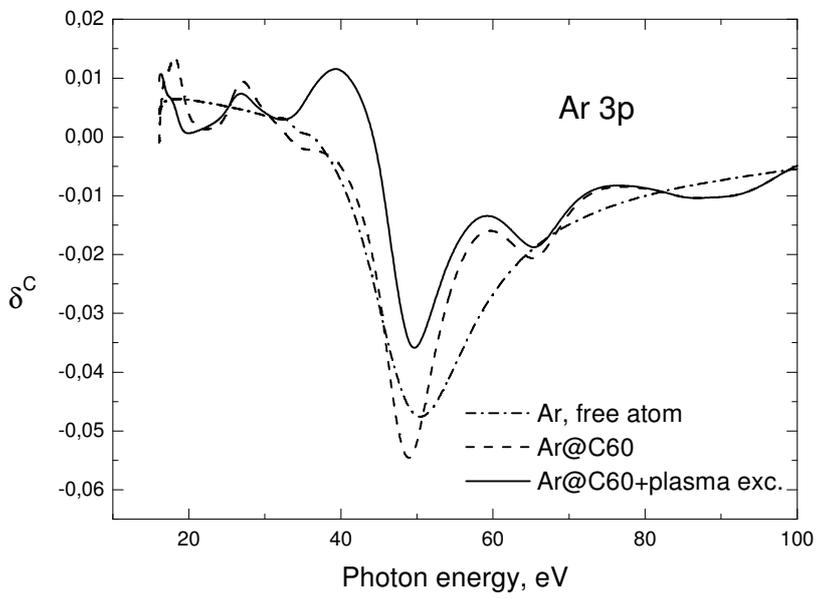

Fig.5. Non-dipole angular anisotropy parameters $\gamma^C$ and $\delta^C$ for 3p electrons of Ar@$C_{60}$

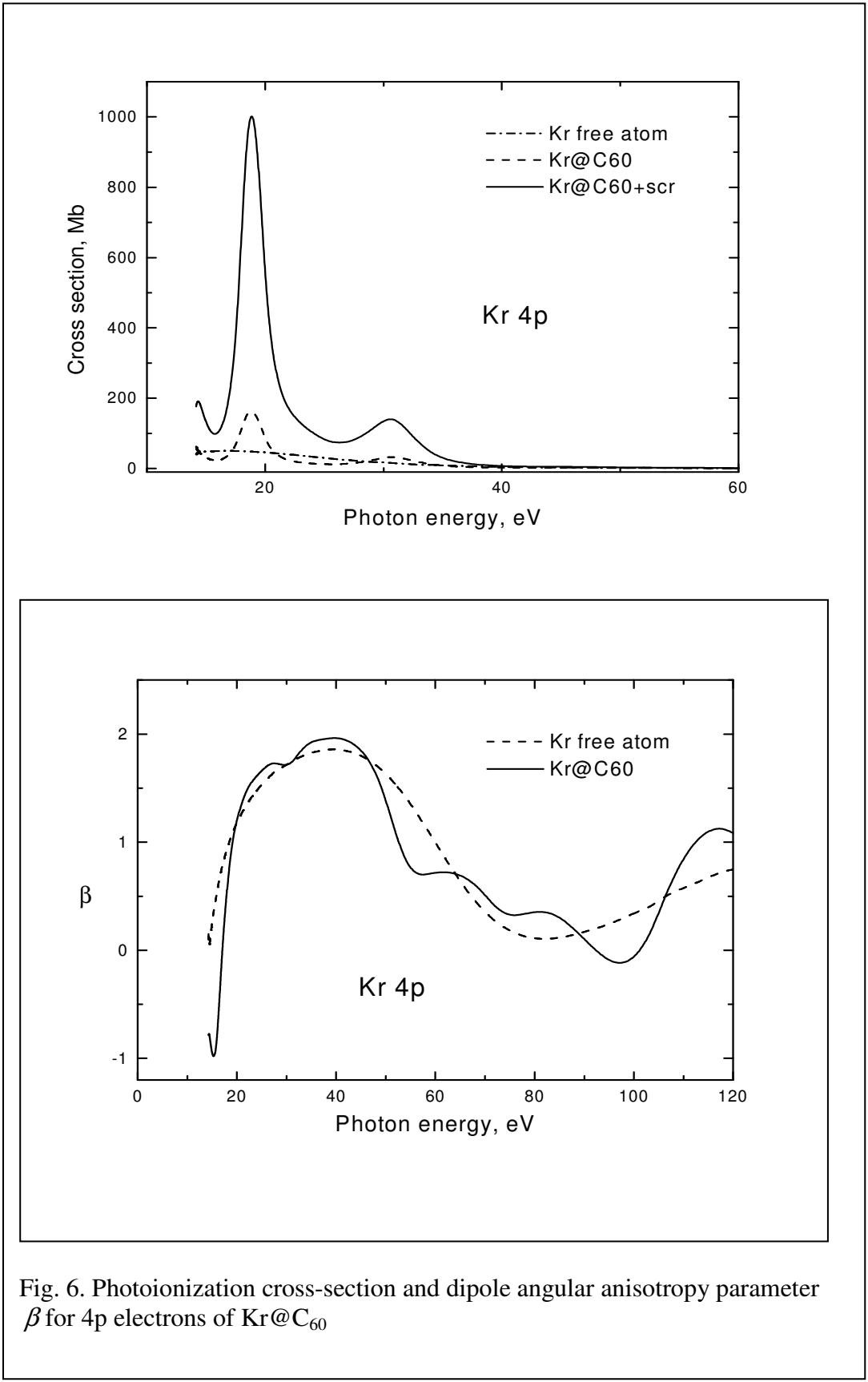

Fig. 6. Photoionization cross-section and dipole angular anisotropy parameter $\beta$ for 4p electrons of $Kr@C_{60}$

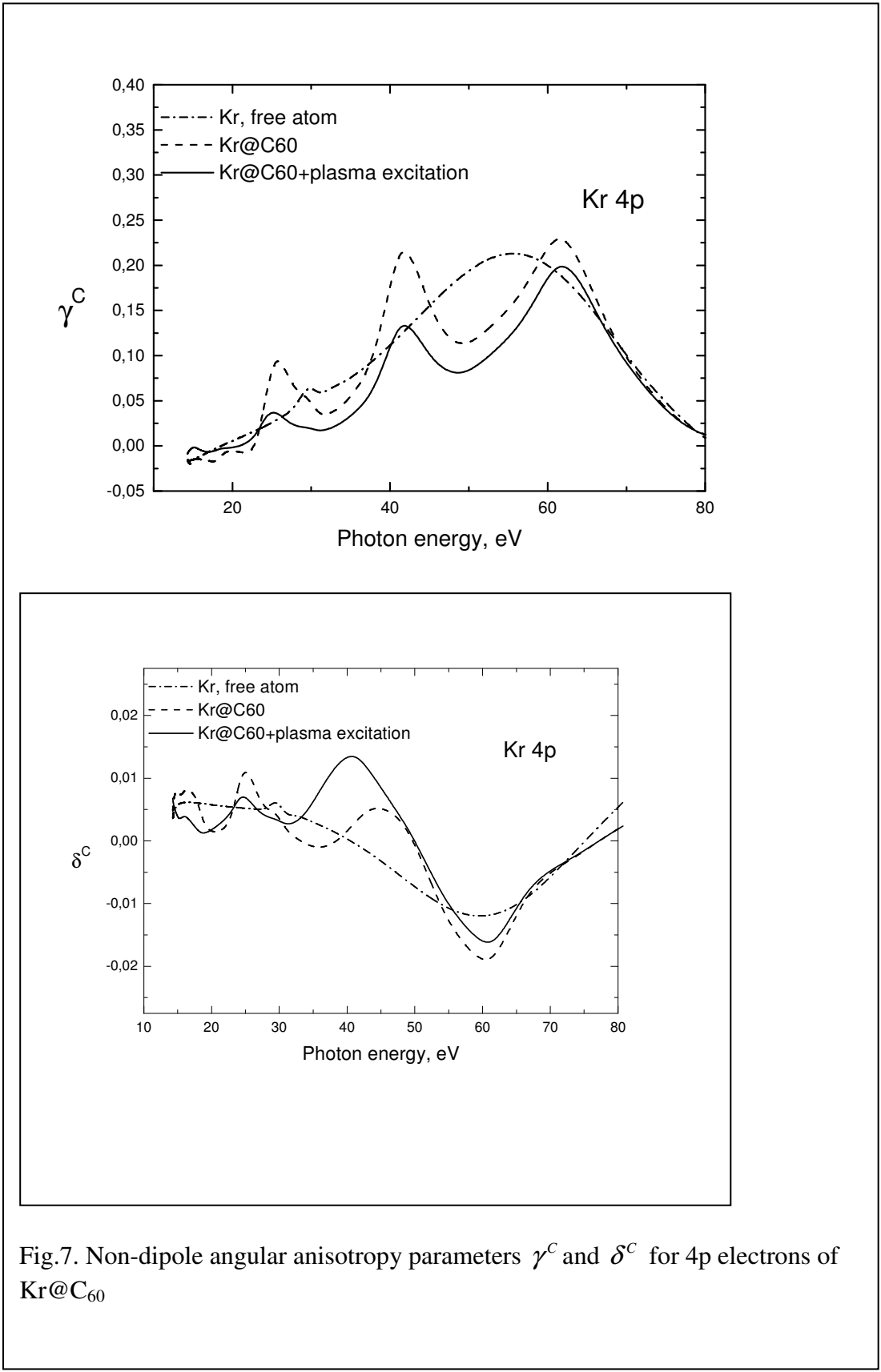

Fig.7. Non-dipole angular anisotropy parameters $\gamma^C$ and $\delta^C$ for 4p electrons of Kr@$C_{60}$

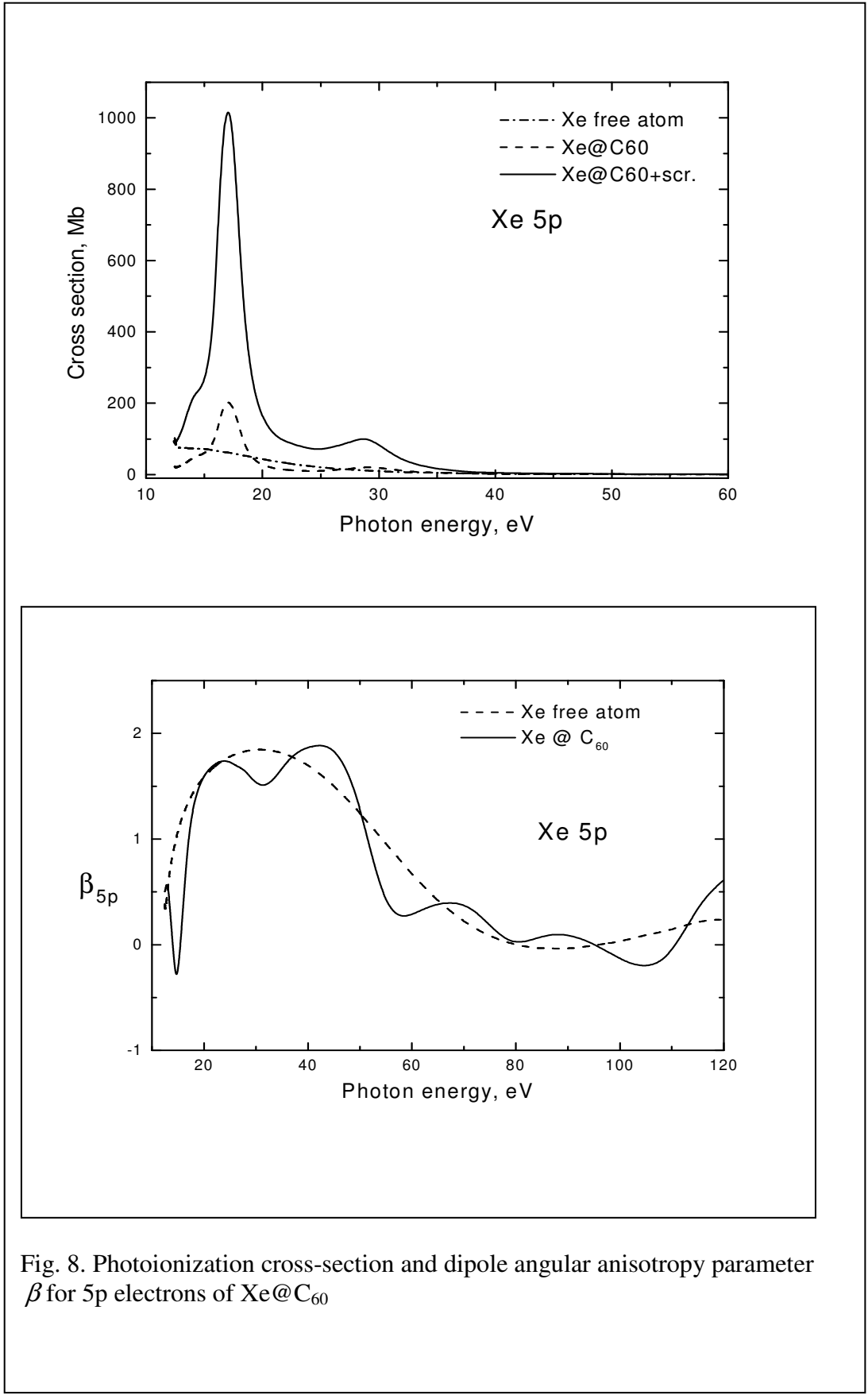

Fig. 8. Photoionization cross-section and dipole angular anisotropy parameter $\beta$ for 5p electrons of Xe@C$_{60}$

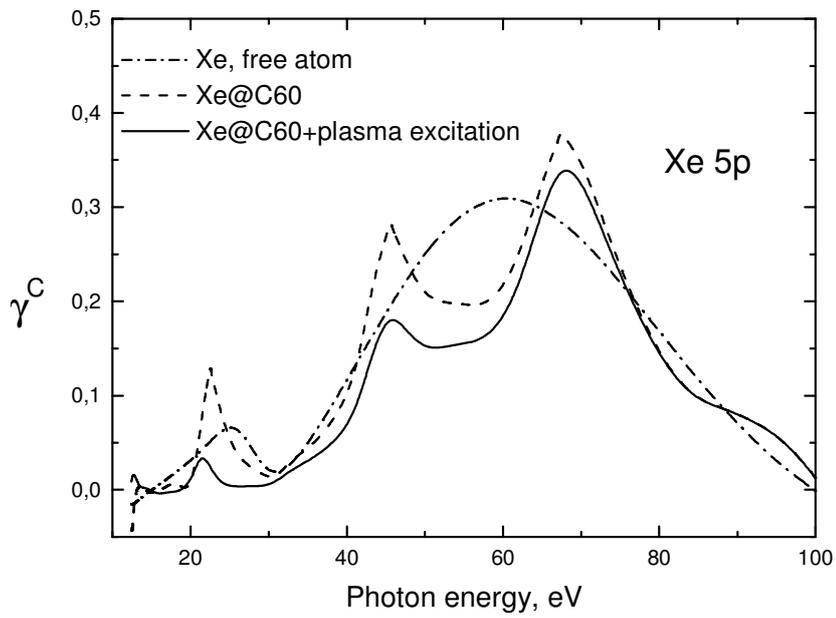

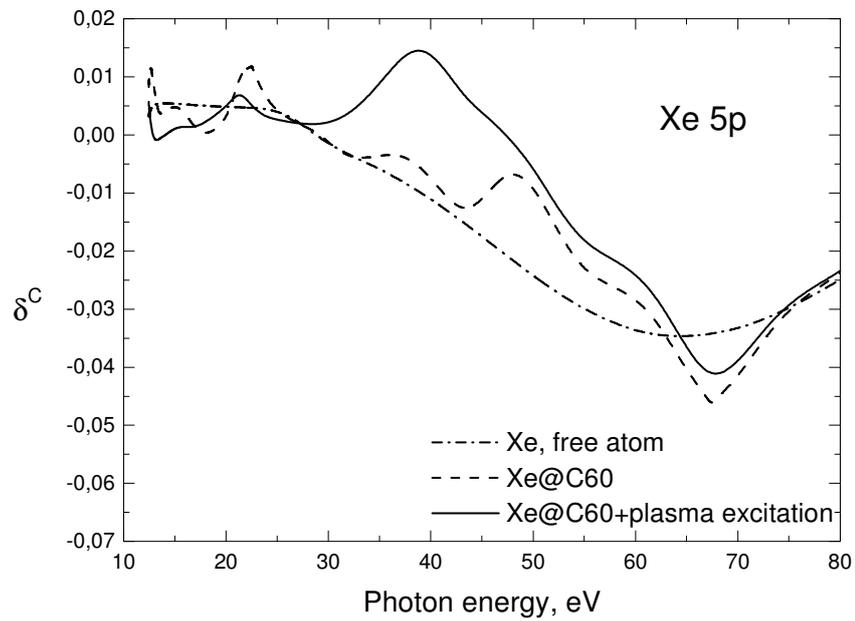

Fig.9. Non-dipole angular anisotropy parameters $\gamma^C$ and $\delta^C$ for 5p electrons of Xe@$C_{60}$